\RequirePackage{fix-cm}
\documentclass[twocolumn]{svjour3}          
\smartqed  
\usepackage[caption=false]{subfig}
\usepackage{graphicx}
\usepackage{amsmath} 
\usepackage{amssymb}
\usepackage{bm}
\usepackage{enumerate} 
\usepackage{multirow} 
\usepackage{csquotes}
\usepackage[usenames,dvipsnames]{color}
\usepackage{cite}
\usepackage[title]{appendix}
\usepackage{times}
\usepackage{latexsym}
\usepackage{xy}
\usepackage{algorithm} 
\newtheorem{theory}{Theorem}
\newtheorem{Rem}{Remark}
\newtheorem{assumption}{Assumption}

\numberwithin{equation}{section}

\usepackage{algcompatible}
\algnewcommand\INPUT{\item[\textbf{Input:}]}
\algnewcommand\OUTPUT{\item[\textbf{Output:}]}

\begin{document}

\title{Regularized Nonlinear Regression for Simultaneously Selecting and Estimating Key Model Parameters
}


\author{Kyubaek Yoon \and Hojun You \and Wei-Ying Wu \and Chae Young Lim \and Jongeun~Choi \and Connor Boss \and Ahmed Ramadan \and John M. Popovich Jr. \and Jacek Cholewicki \and N.~ Peter Reeves \and Clark J. Radcliffe 
}


\institute{Kyubaek Yoon \and Jongeun Choi (Corresponding author) \at
              The School of Mechanical Engineering, Yonsei University, 50 Yonsei Ro, Seodaemun Gu, Seoul 03722, Republic of Korea \\
              \email{jongeunchoi@yonsei.ac.kr }           
           \and
           Hojun You \and Chae Young Lim \at
              Department of Statistics, Seoul National University, Seoul 08826, Republic of Korea
        \and
            Wei-Ying Wu \at
              Department of Applied Mathematics, National Dong Hwa University, Hualien 97401, Taiwan
        \and
            Connor Boss \at
              Department of Electrical Engineering, Michigan State University, East Lansing, MI 48824, USA
        \and
            Ahmed Ramadan \at
              Department of Physical Therapy and Rehabilitation Science, University of Maryland, Baltimore, MD 21201, USA
        \and
            John M. Popovich Jr. \and Jacek Cholewicki  \at
              MSU Center for Orthopedic Research, Department of Osteopathic Surgical Specialties, Michigan State University, East Lansing, MI 48824 USA
        \and
            N. Peter Reeves \at
              Sumaq Life LLC, East Lansing, MI 48823 USA
        \and
            Clark J. Radcliffe \at
              Department of Mechanical Engineering and MSU Center for Orthopedic Research, Michigan State University, East Lansing, MI 48824 USA
}

\date{Received: date / Accepted: date}

\maketitle

\begin{abstract}
\sloppy
In system identification, estimating parameters of a model using limited observations results in poor identifiability. To cope with this issue, we propose a new method to simultaneously select and estimate sensitive parameters as key model parameters and fix the remaining parameters to a set of typical values. Our method is formulated as a nonlinear least squares estimator with $L_{1}$-regularization on the deviation of parameters from a set of typical values. First, we provide consistency and oracle properties of the proposed estimator as a theoretical foundation. Second, we provide a novel approach based on Levenberg-Marquardt optimization to numerically find the solution to the formulated problem. Third, to show the effectiveness, we present an application identifying a biomechanical parametric model of a head position tracking task for 10 human subjects from limited data. In a simulation study, we analyze the bias and variance of estimated parameters. In an experimental study, our method improves the model interpretation by reducing the number of parameters to be estimated while maintaining variance accounted for (VAF) at above 82.5$\%$. Moreover, the variance of estimated parameters is reduced by 71.1$\%$ as compared to that of the estimated parameters without ${L}_{1}$-regularization. Our method is 54 times faster than the standard simplex-based optimization to solve the regularized nonlinear regression.
\end{abstract}

\section{Introduction}
\label{introduction}
In parameter estimation, a model is considered to be identifiable when a unique set of parameters is specified for given measurement data. 
However, when the data is limited, estimating unknown parameters of a model results in poor identifiability \cite{gutenkunst2007universally}. In such a case, small changes in the data could result in very different
estimated parameters, for example, a rather randomly chosen local minimum out of multiple local minima \cite{lund2019global,ramadan2018selecting}. 
The resulted overfitting impairs the model parsimony and generalizability \cite{pitt2002good}. The overfitted model may yield good results with a training data set  used to estimate parameters, but it may yield poor estimates with a new test data set.
Moreover, this issue becomes worse when parameters are estimated from the data corrupted by random noise \cite{geman1992neural}.

A biomechanical model often has unknown parameters to be estimated with limited data due to unavailable internal states and the non-invasive nature of human data collection \cite{little2010parameter}.
For a limited observation data set, one way for improving identifiability is to build a parsimonious model by lumping a large number of parameters into a small number of lumped parameters \cite{do2018appearance}. Such a parsimonious model has better interpretability and provides   higher estimation accuracy for arbitrary data \cite{babyak2004you}. 
 
As a way to build a parsimonious model, the Least absolute shrinkage and selection operator (Lasso) was first introduced in \cite{tibshirani1996regression} and then further developed in \cite{zou2005regularization,tibshirani2005sparsity,yuan2006model,zou2006adaptive}. The Lasso is typically used to select sensitive parameters among parameters of a linear model. A sensitive parameter subset is considered to have relatively large impact on the output of the model. That is, small changes in the sensitive parameters result in large changes in the model response \cite{tibshirani1988sensitive}. 

To formally introduce the Lasso, we suppose that we have $(\bm x_{i},y_i), i \in \{1,\cdots,n\}$ where $\bm x_i = [x_i^{[1]},\cdots,x_i^{[p]}]$ and $y_{i}=f({\bm x_{i}};\bm\theta_0)+\epsilon_{i}$. $f({\bm x_{i}};\bm\theta_0)$ is a function, which depends on the true parameter vector $\bm\theta_0$. $\epsilon_{i}$ is independent and identically distributed with $\mathbb{E}(\epsilon)=0$ and $Var(\epsilon_i)=\sigma^2$. Without loss of generality, we   assume   that the true parameter vector $\bm\theta_0=[\theta_{01},\theta_{02},...,\theta_{0s},\theta_{0s+1},...,\theta_{0p}]^T$ has the first $s$ entries non-zero. That is, $\theta_{0k}\neq 0$, for $1\leq k\leq s$
and $\theta_{0k}= 0$, for $s+1 \leq k\leq p$.
Finally, when $f({\bm x_{i}};\bm\theta_0)={\bm x_{i}\bm\theta_0}$, we consider the following linear least squares problem with $L_{1}$-regularization.
\begin{equation}\label{eq6}
\begin{aligned}
   \hat{\bm\theta} &= \arg\min\limits_{\theta} \left [ \| \bm y-\bm{X}\bm\theta\|^{2}_{2} +n \lambda  \sum_{k=1}^{p}|\theta_k|  \right ],
\end{aligned}
\end{equation}
where $\bm X = [\bm x_1,\bm x_2,\cdots, \bm x_n] \in \mathbb{R}^{n\times p}$ is the input. $\bm y = [y_1, y_2,\cdots, y_n]^T \in \mathbb{R}^{n\times 1}$ is the   observation vector. In~\eqref{eq6},  the hyperparameter  $\lambda > 0\in \mathbb{R}$  determines the amount of regularization. The Lasso shrinks more number of parameters toward 0 as $\lambda$ increases in general. Moreover, insensitive parameters are shrunk to 0 if $\lambda$ is sufficiently large \cite{zou2006adaptive}. The remaining parameters, which are not shrunk to 0, are considered sensitive parameters \cite{ramadan2018selecting}. In this paper, we consider the sensitive parameters as key model parameters.
The ${L}_{1}$-regularization methods and weighted least squares method were used  to select nonlinear auto regressive with exogenous variables (NARX) models~\cite{qin2012selection}. The Lasso was used to remove insensitive parameters when all unknown parameters are to be non-negative~\cite{kump2012variable}. The modified Lasso was proposed  for the nonlinear induction motor identification problem, which deals with a similar problem to our study~\cite{rasouli2012reducing}. However, they do not provide consistency and the asymptotic normality results. The modified principal component analysis (PCA) based on the Lasso was proposed  for dimension reduction~\cite{jolliffe2003modified}. To select and estimate sensitive parameters of a model, sensitive parameters were selected
based on parameter estimation variances predicted by the Fisher information
matrix \cite{ramadan2018selecting,ramadan2019feasibility}. 

In this paper, our objective is to develop a regularized nonlinear parameter estimation method for a model with unknown parameters using a limited data set. 
Thus, we formulate the model parameter estimation problem as a nonlinear least squares problem with $L_{1}$-regularization as follows.
\begin{equation} \label{ourform}
\begin{aligned}
 \hat { \tilde  { \bm\theta} } =& \arg\min\limits_{{\tilde{\bm\theta}}} \left [ \| \bm y-{f}(\bm{X}; \tilde {\bm\theta})\|^{2}_{2} + n \lambda\sum_{k=1}^{p}| \tilde \theta_k|  \right ],
\end{aligned}
\end{equation}
where ${\tilde{\bm\theta}}:=\bm{\theta}-{\bar {\bm\theta}}$ and ${\bar {\bm\theta}}$ is the set of typical values. Note that these values may be obtained as the mean values of $\bm{\theta}$ based on preliminary information or estimation. The details of (\ref{ourform}) are introduced in Section~\ref{application}.
In our preliminary work, we developed a parameter selection method for system identification with application to head-neck position tracking and reported the model parameter estimates \cite{kyubaek2019penalized}.

\sloppy
The contributions of the paper are as follows. First,  we consider  nonlinear regression with a generalized penalty function that includes an ${L}_{1}$-penalty function and provide its  consistency and oracle properties (i.e., convergence to the correct sparsity and asymptotic normality) in Section~\ref{theory}. To the best of the authors' knowledge, our work is the first to provide such analysis for nonlinear regression   with a generalized penalty function. For example, convergence properties for various penalized linear regression have been discussed \cite{johnson2008penalized,fan2001scad}.
Note that we do not assume the distribution of errors, which is different from the assumption of \cite{fan2001scad}.
In Section~\ref{application}, we then reformulate the regularized nonlinear regression for simultaneously selecting and estimating key model parameters by defining $\tilde \theta_k$ in \eqref{ourform} as the deviation of the k-th parameter from its  nominal value. Next, we improve the optimization algorithm of \cite{rasouli2012reducing} to numerically solve the nonlinear least squares problem.
Finally, to show the effectiveness, we present an application identifying a biomechanical parametric model of a head-neck position tracking task from limited data in simulation and experimental studies. In a simulation study, our algorithm reduces the variance of most parameter estimates as well as the bias. In an experimental study, the variance of selected sensitive parameters is reduced by $71.1\%$ on average while maintaining the goodness of fit at above $82.5\%$. In addition, our method is 54 times faster as compared to the Lasso with the brute force optimization, e.g., the standard simplex-based optimization we presented recently \cite{ramadan2018selecting}.

\section{Consistency and Oracle Properties} \label{theory}
We propose a penalized nonlinear regression approach for parameter selection and estimation. First of all, we adopt the following equivalent nonlinear least squares estimator with a generalized penalty function from (\ref{ourform}):
\begin{equation}\label{eq:prob}
    \hat{\bm\theta}_n=\arg \min_{\bm\theta\in D} \left [\bm Q_n(\bm\theta):=\bm S_n(\bm\theta)+n\sum^{p}_{k=1}p_{\lambda_n}(|\theta_k|) \right ],
    \tag{3}
\end{equation}
where 
$\bm S_n(\bm\theta)=\sum^{n}_{i=1}\left(y_i-f({\bm{x}_i};\bm\theta)\right)^2$. 
The first term in $\bm Q_n(\cdot)$ corresponds to nonlinear least squares estimation, and 
the second term, $p_{\lambda_n}(\cdot)$ is the penalty function used for parameter selection. $\lambda_n$ in the penalty function is a nonnegative regularization parameter. Note that we consider a generalized penalty function in~\eqref{eq:prob}. The appropriate choice  of the penalty function, including $L_{1}$-regularization, is further investigated in the assumption ~\ref{eq:penalty:assumption}.

Without loss of generality, we assume only a few parameters are non-zero, such that the true parameter vector $\bm\theta_0=[\theta_{01},\theta_{02},...,\theta_{0s},\theta_{0s+1},...,\theta_{0p}]$ has the first $s$ entries non-zero. That is, $\theta_{0k}\neq 0$, for $1\leq k\leq s$
and $\theta_{0k}= 0$, for $s+1 \leq k\leq p$. For the nonlinear function $f(\cdot;\cdot)$ in~\eqref{eq:prob}, we further consider the following assumptions.

\begin{assumption}\label{eq:nonlinear:assumption}(Nonlinear function)
	\begin{enumerate}
		\item The true parameter $\bm\theta_0$ is in the interior of the bounded parameter set $\Theta$, and $f({\bm{x}_i};\bm\theta)$ is twice differentiable with respect to $\bm\theta$ near $\bm\theta_0$ for all $i$.
		\item Let $\bm{f}_k(\bm\theta)=\left(\frac{\partial{f(\bm x_1,\bm\theta)}}{\partial\theta_k},\ldots,	\frac{\partial f(\bm x_n,\bm \theta)}{\partial\theta_k}\right)^{T}$ and  ${\bm{\dot{F}}}(\bm\theta)=(\bm{f}_1,...,\bm{f}_p)$. Then, there exists a positive definite matrix $\Gamma$ such that $\frac{1}{n} {\bm{\dot{F}}}(\bm\theta_0)^{T}{\bm{\dot{F}}}(\bm\theta_0)\rightarrow \Gamma$ as $n\rightarrow \infty$. 
		\item As $n\rightarrow \infty$ and $\|\bm\theta_1-\bm\theta_0\|\rightarrow 0$,  $$ {\bm{\dot{F}}}(\bm \theta_1)^{T} {\bm{\dot{F}}}(\bm\theta_1)\left({\bm{\dot{F}}}(\bm\theta_0)^{T} {\bm{\dot{F}}}(\bm\theta_0)\right)^{-1}  \rightarrow I$$ 
		uniformly, where $I$ is the identity matrix
		\item There exists a $\delta>0$ such that $$\limsup_{n\rightarrow \infty}\frac{1}{n}\sum^{n}_{i=1}\sup_{\|\bm\theta-\bm\theta_0\|\leq \delta}\left(\frac{\partial^2 f(\bm{x}_i;\bm\theta)}{\partial\theta_k \partial\theta_s}\right)^2<\infty$$
	\end{enumerate}
\end{assumption}
For the penalty function $p_{\lambda_n}(\cdot)$ in~\eqref{eq:prob}, we further consider the following assumptions.

\begin{assumption}\label{eq:penalty:assumption}(Penalty function)
	The first and second derivative of the penalty function $p_{\lambda_n}(\cdot)$ denoted by $q_{\lambda_n}(\cdot)$ and $q^{\prime}_{\lambda_n}(\cdot)$ have the following properties:
	\begin{enumerate}	
		\item For a nonzero fixed $\theta$, \[\lim_{n \to \infty}  n^{1/2}q_{\lambda_n}(|\theta|)=0, \lim_{n \to \infty} q^{\prime}_{\lambda_n}(|\theta|)=0.\]
		\item For any $M>0$, \[\lim_{n \to \infty} n^{1/2}\inf_{|\theta|\leq Mn^{-1/2}} q_{\lambda_n}(|\theta|)\rightarrow \infty.\]
	\end{enumerate}
\end{assumption}

\begin{Rem}
	Assumption~\ref{eq:penalty:assumption} is satisfied for several well known penalty functions, e.g., SCAD, Adaptive Lasso, and Hard penalty with proper choices of $\lambda_n$. Assumption~\ref{eq:penalty:assumption}-(1) is satisfied for $L_{1}$-regularization with a proper choice of $\lambda_n$. The details are discussed in \cite{johnson2008penalized}.
\end{Rem}

The following theorem shows the existence of a local minimizer of $\bm Q_n(\bm \theta)$ with the order of $O_p(n^{-1/2})$.
\begin{lemma}\label{lem:Consistency}
	Under Assumptions~\ref{eq:nonlinear:assumption} and~\ref{eq:penalty:assumption}-(1), for any  $\eta>0$, there exists a positive constant $C$ that makes, for large enough n,
	$$P\left(\inf_{\|\bm v\|=C} \bm Q_n(\bm\theta_0+n^{-1/2} \bm v)-\bm Q_n(\bm\theta_0)>0\right)>1-\eta,$$ where $\bm Q_n(\bm\theta)=\bm S_n(\bm\theta)+n\sum^{p}_{k=1}p_{\lambda_n}(|\theta_k|)$.
\end{lemma}
\vspace*{0.2cm}
\begin{theory}\label{thm:Consistency}
	Under the assumptions in Lemma~\ref{lem:Consistency}, there exists, with probability tending to 1, a root-n-consistent local minimizer $\hat{\bm\theta}$ of $\bm Q_n(\bm\theta)$, that is, $\|\hat{\bm\theta} -\bm\theta_0\| = O_p(n^{-1/2})$.
\end{theory}

Next theorem shows oracle properties (i.e., convergence to the correct sparsity and asymptotic normality) of the estimator on the true set.

\begin{theory}\label{thm:sparsity}
	Assume that $\hat{\bm\theta}=(\hat{\theta}_k)^p_{k=1}$ is the local minimizer of $\bm Q_n(\bm\theta)$ with the root-n-consistency. 
	If Assumptions~\ref{eq:nonlinear:assumption} and~\ref{eq:penalty:assumption} hold, 
	\begin{itemize}
		\item[(i)] for the set $M_k=\{\omega:\hat{\theta}_k\neq 0\}$, ~$s+1\leq k \leq p$,~ $$P(M_k)~\rightarrow~ 0$$
		\item[(ii)] For $\hat{\bm\theta}_{11}=(\hat{\theta}_{1},\hat{\theta}_{2},...,\hat{\theta}_{s})^{T}$, ${\bm\theta}_{01}=({\theta}_{01},{\theta}_{02},...,{\theta}_{0s})^{T}$, $$n^{1/2}(2\Gamma_{11})({\hat{\bm \theta}}_{11}-\bm\theta_{01}+(2\Gamma_{11})^{-1}b_n)~\stackrel{d}{\longrightarrow}~ N(0,2\Gamma_{11}\sigma^2),$$ where  $b_n=(q_{\lambda_n}({|\theta_{01}|)\text{sgn}(\theta}_{01}),q_{\lambda_n}({|\theta_{02}|)\text{sgn}(\theta}_{02}),...$, $q_{\lambda_n}(|\theta_{0s}|)\text{sgn}(\theta_{0s}))^{T}$ and $\Gamma_{11}$ is the first $s\times s$ submatrix of $\Gamma$.
	\end{itemize}
\end{theory}

The proofs of Lemma~\ref{lem:Consistency} and Theorem~\ref{thm:Consistency} are given in Appendix~\ref{A_1} and Appendix~\ref{A_2}, respectively and the proof of Theorem~\ref{thm:sparsity} is given  in Appendix~\ref{A_3}.\\
\begin{figure*}[t]
    \centering
    \includegraphics[width=17cm]{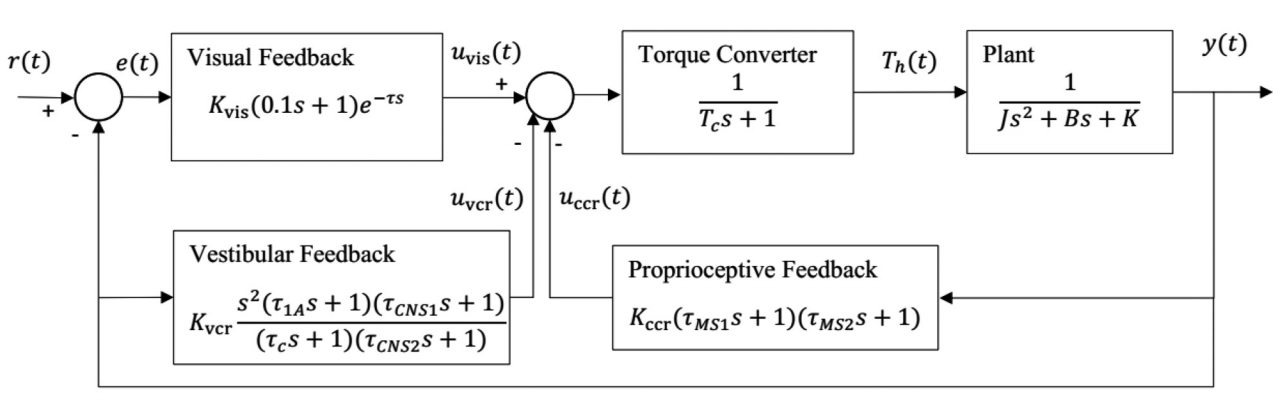}
    \caption{Sensorimotor control block diagram for the head-neck system~\cite{ramadan2018selecting}}
    \label{fig1}
\end{figure*}
\section{Application to  Head  Position Tracking}\label{application}
In this section, we evaluate our approach to solve the nonlinear least squares problem with $L_1$-regularization from simulation and experimental studies of a biomechanical parametric model of a head-neck position tracking task in \cite{ramadan2018selecting}. The reliability of the head-neck position tracking task to quantify head-neck motor control is demonstrated in \cite{popovich2015quantitative}.
As compared to the Levenberg optimization algorithm in \cite{rasouli2012reducing}, our method implements the Levenberg-Marquardt optimization algorithm to numerically solve our nonlinear least squares problem with $L_1$-regularization. The Levenberg-Marquardt optimization uses the diagonal elements of the hessian matrix approximation to overcome the slow convergence problem when the value of the damping factor is large \cite{more1978levenberg}. The details of our algorithm are shown in Appendix~\ref{ours} and Table~\ref{alg:lasso}.

Additionally, in order to simultaneously select and estimate key model parameters of the head-neck system, we reformulate the Lasso penalty function as ${L}_{1}$-regularization on the deviation of parameters from a set of typical values. In this paper, we adopt the mean of the nominal parameter values obtained from preliminary estimation as the set of typical values. Therefore, our method simultaneously selects and estimates only sensitive parameters while fixing insensitive parameters onto the mean of the nominal parameter values. In order to compare with the method of  \cite{ramadan2018selecting}, we set the number of sensitive parameters to 5. In this case, we increase the regularization hyperparameter value until we obtain 5 sensitive parameters. The goodness of fit is quantitatively evaluated by variance accounted for (VAF). VAF represents how much the experimental data can be explained by a fitted regression model. VAF is formally defined as follows.
\begin{equation*}
    {\text{VAF}}(\bm\theta)(\%) = \left [1 - \frac{\sum^{n}_{i=1}(y_i-\hat{y}_i(\bm\theta))^{2}}{\sum^{n}_{i=1}y_i^{2}} \right ] \times 100
\end{equation*}

\begin{table*}[t]
\centering
\caption{The neurophysiological parameters of the head position model. All the information is obtained from \cite{ramadan2018selecting} }
\label{Table 1}
\begin{tabular}{c c c c l }
\hline\hline
\multicolumn{2}{c}{\textbf{Parameters}}& \multicolumn{1}{c}{\textbf{Max}}&
\multicolumn{1}{c}{\textbf{Min}}&
\multicolumn{1}{c}{\textbf{Description}}\\ 
\hline
\multirow{12}{*}{$\bm{\theta}$}&$K_{vis}[\frac{Nm}{rad}]$&$10^{3}$&50&Visual feedback gain\\

    &$K_{vcr}[\frac{Nms^2}{rad}]$&$10^{4}$&500&Vestibular feedback gain\\

    &$K_{ccr}[\frac{Nm}{rad}]$&300&1&Proprioceptive feedback gain\\

    &$\tau[s]$&0.4&0.1&Visual feedback delay\\

    &$\tau_{1A}[s]$&0.2&0.01&Lead time constant of the irregular vestibular afferent neurons\\

    &$\tau_{CNS1}[s]$&1&0.05&Lead time constant of the central nervous system\\

    &$\tau_{C}[s]$&5&0.1&Lag time constant of the irregular vestibular afferent neurons\\

    &$\tau_{CNS2}[s]$&60&5&Lag time constant of the central nervous system\\

    &$\tau_{MS1}[s]$&1&0.01&First lead time constant of the neck muscle spindle\\

    &$\tau_{MS2}[s]$&1&0.01&Second lead time constant of the neck muscle spindle\\

    &$B[\frac{Nm}{rad}]$&5&0.1&Intrinsic damping\\

    &$K[\frac{Nm}{rad}]$&5&0.1&Intrinsic stiffness\\
\hline

\multirow{2}{*}{$\bm{\theta_{fixed}}$}&$J[kgm^2]$&0.0148&0.0148&Head inertia\\

    &$T_{c}[s]$&0.1&0.1&Torque converter time constant\\

\hline\hline
\end{tabular}
\end{table*}

\subsection{Subjects}
10 healthy subjects participated in the experimental study. They did not have any history of neck pain lasting more than three days or any neurological motor control impairments. The Michigan State University’s Biomedical and Health Institutional Review Board approved the test protocol. The subjects signed an informed consent before participating in the experiment \cite{ramadan2018selecting,ramadan2019feasibility}.

\begin{figure}[t!]
    \centering
    \includegraphics[width=8cm]{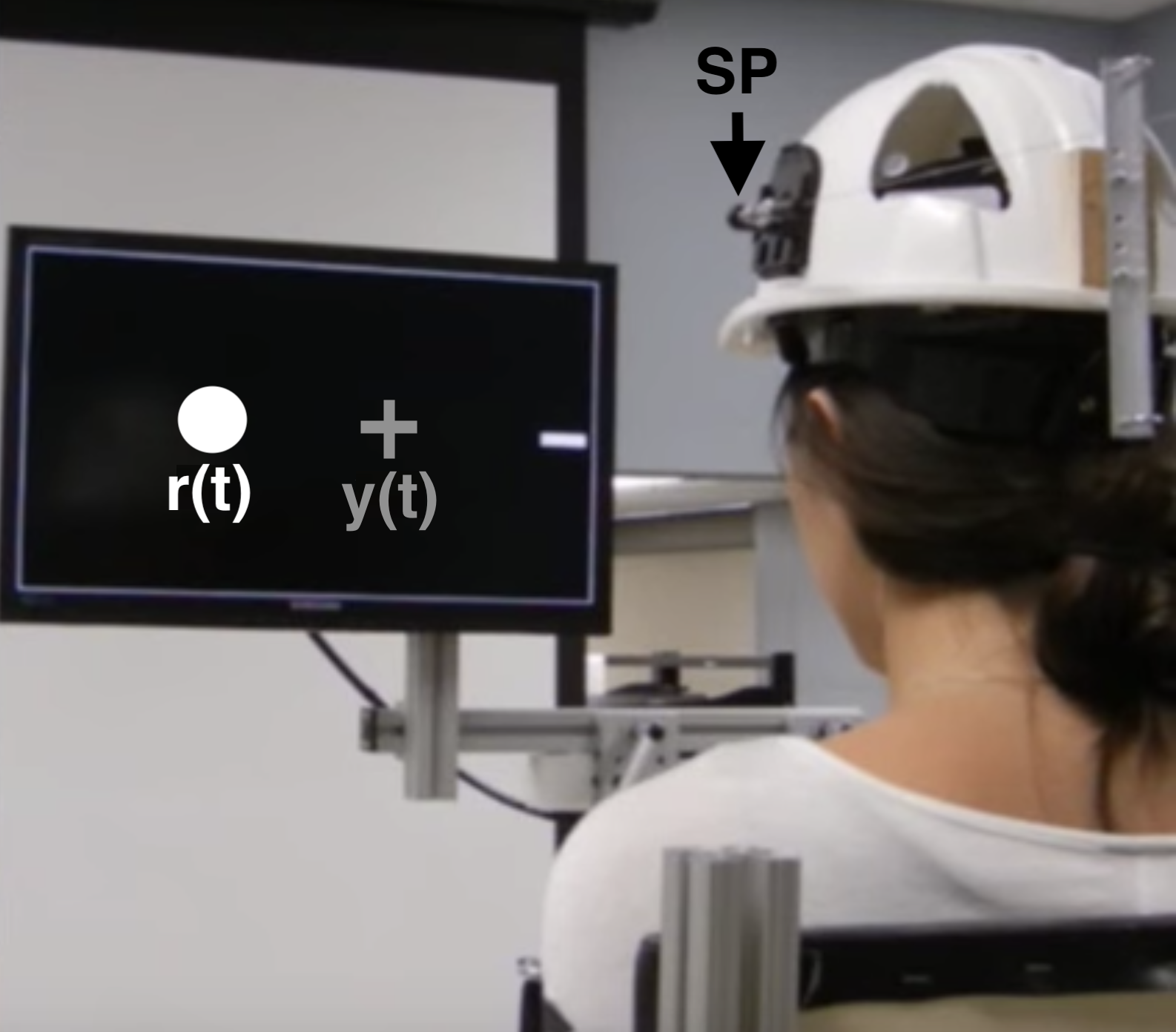}
    \caption{The experimental setup for the head-neck position tracking task. $r(t)$ is the reference command signal, and $y(t)$ is the measured head rotation angle, and SP indicates one of string potentiometers on both sides of the helmet.}
    \label{fig2}
\end{figure} 

\subsection{Parametric model}
Fig.~\ref{fig1} shows the block diagram of the head-neck system for  position tracking. This is a representative physiological feedback control model~\cite{peng1996dynamical,chen2002modeling}.
The model consists of 14 parameters.  As shown in Table~\ref{Table 1}, 2  out of 14 parameters are set to fixed values from \cite{peng1996dynamical}.  The remaining 12 parameters to be estimated are:
\begin{equation*}
\begin{split}
        \boldsymbol{\theta} = [{K}_{vis} \ {K}_{vcr} \ {K}_{ccr} \ \tau \ \tau_{1A} \ \tau_{CNS1} \\ \tau_{C} \ \tau_{CNS2} \ \tau_{MS1} \ \tau_{MS2} \ {B} \ {K}]
\end{split}
\end{equation*}
The remaining 12 parameters have the lower and upper bounds from \cite{ramadan2018selecting} and are normalized using min-max normalization in order to ignore the scale differences between parameters.


\begin{figure*}
\centering
\subfloat[]{\includegraphics[width=1\textwidth]{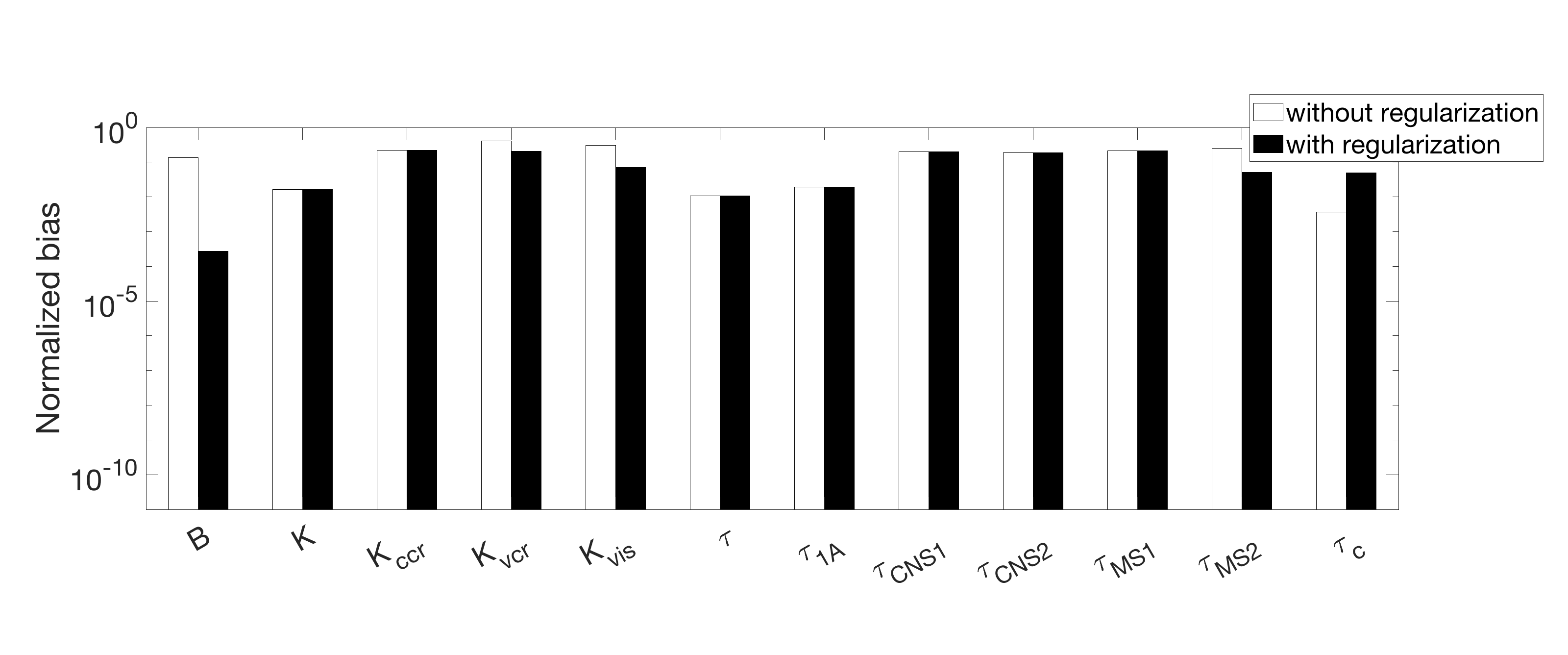}}
\qquad
\subfloat[]{\includegraphics[width=1\textwidth]{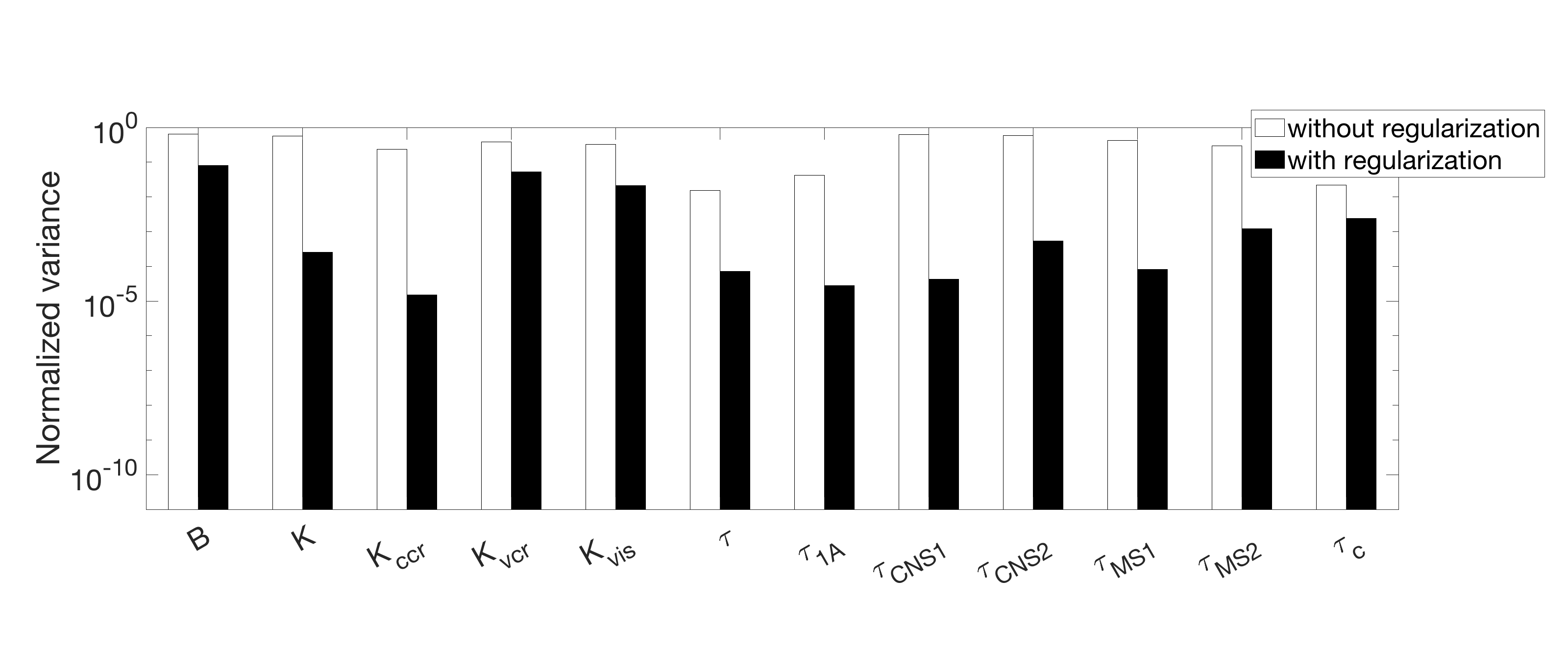}}
\qquad
\caption{The bias (a) and variance (b) of estimated parameters in the simulation studies. Comparing our method with a nonlinear least squares problem without ${L}_{1}$-regularization (white bar), we set the mean of nominal parameters (black bar) obtained from preliminary estimation as a set of typical values. The y-axis is a logarithmic scale.}
\label{bv}
\end{figure*}


\begin{figure*}
\centering
\subfloat[Subject 1]{\label{fig:01}\includegraphics[width=0.95\textwidth]{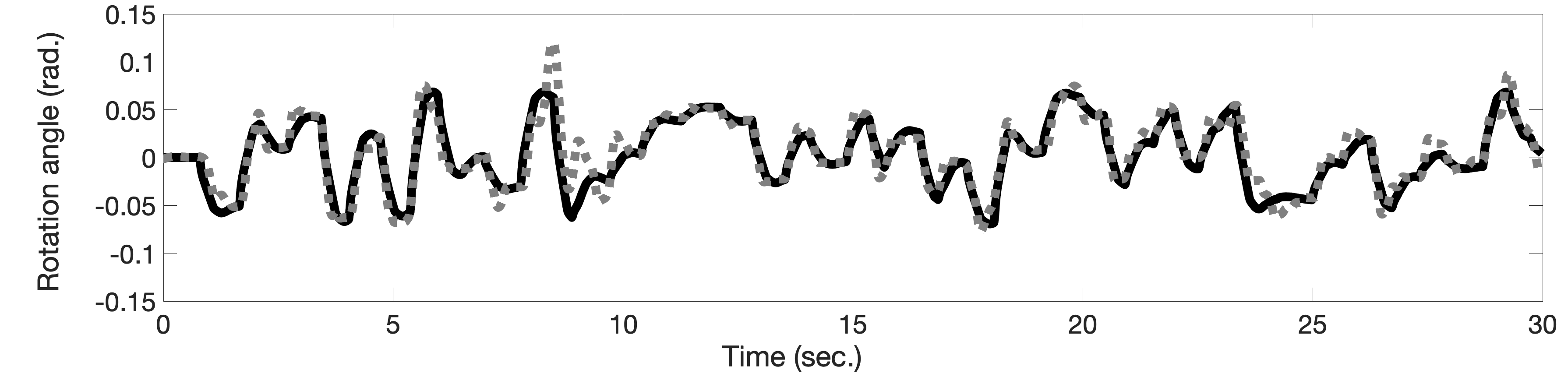}}
\qquad
\subfloat[Subject 2]{\label{fig:02}\includegraphics[width=0.95\textwidth]{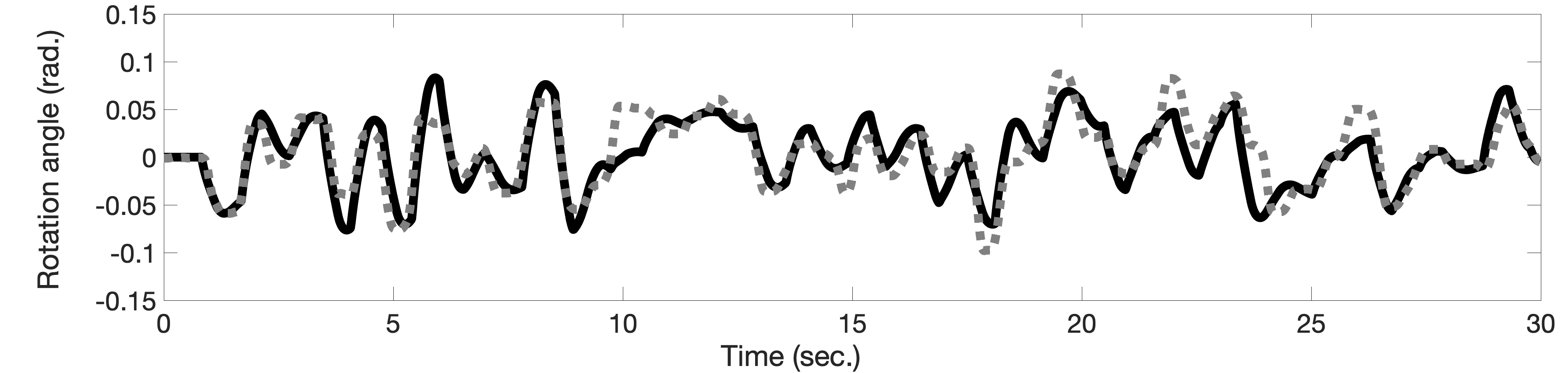}}
\qquad
\subfloat[Subject 3]{\label{fig:03}\includegraphics[width=0.95\textwidth]{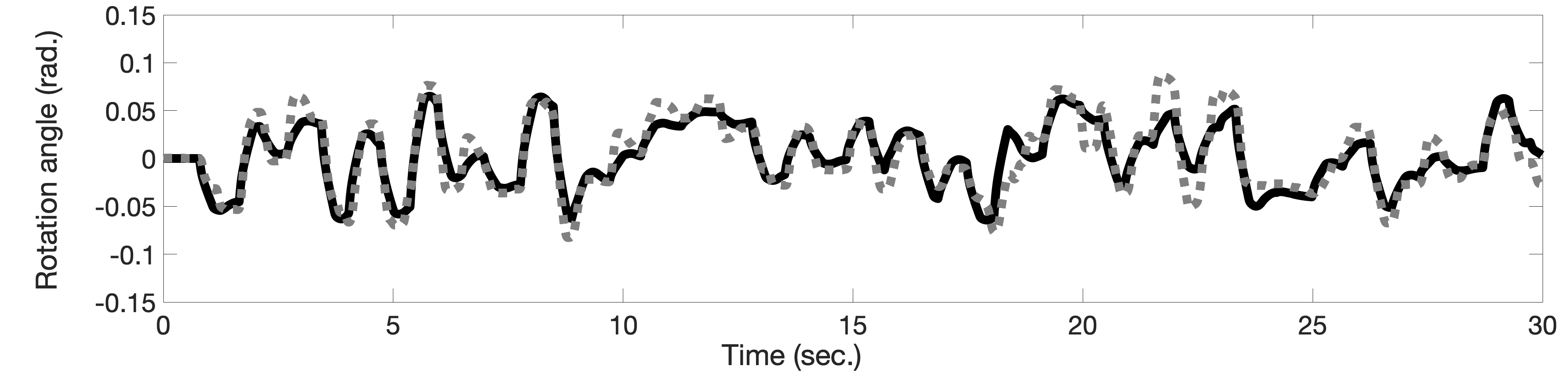}}
\qquad
\subfloat[Subject 4]{\label{fig:04}\includegraphics[width=0.95\textwidth]{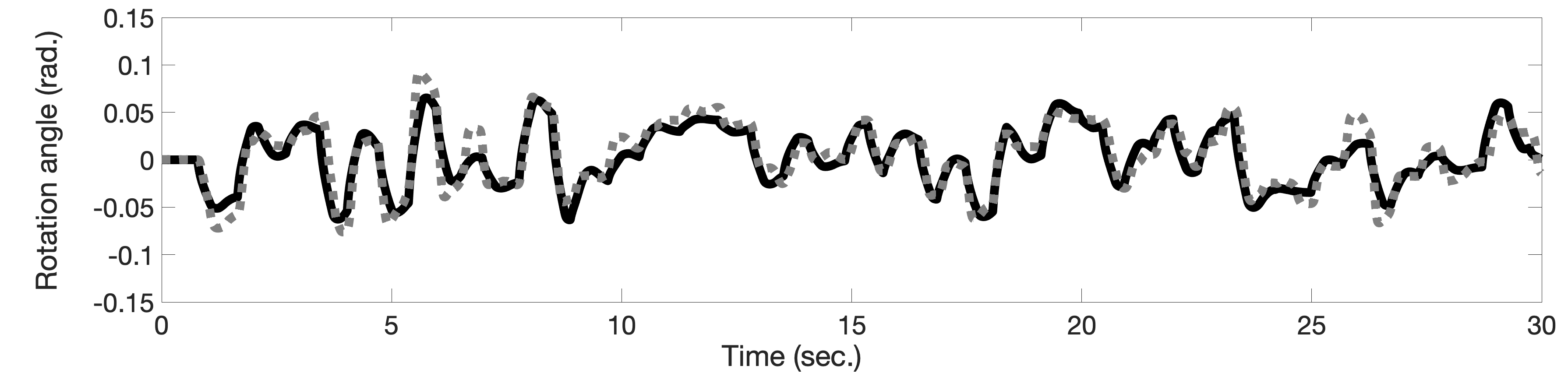}}
\qquad
\subfloat[Subject 5]{\label{fig:05}\includegraphics[width=0.95\textwidth]{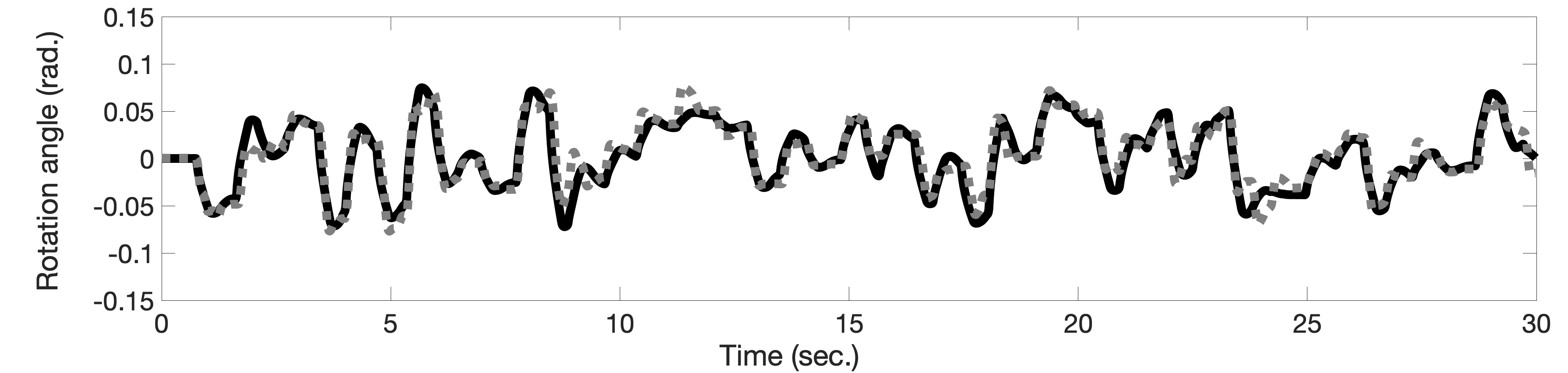}}
\qquad
\caption{The curve fitting with 10 experimental cases. Solid lines represent estimated responses from the fitted models and dotted lines represent measured responses.}
\label{fig:cont1}
\end{figure*}

\begin{figure*}
\ContinuedFloat
\centering
\subfloat[Subject 6]{\label{fig:06}\includegraphics[width=0.95\textwidth]{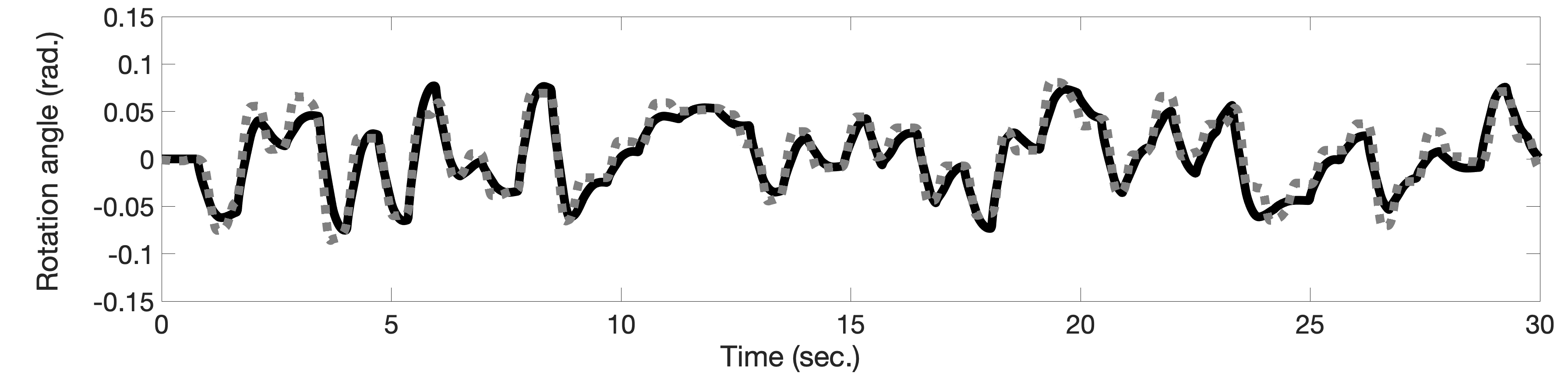}}
\qquad
\subfloat[Subject 7]{\label{fig:07}\includegraphics[width=0.95\textwidth]{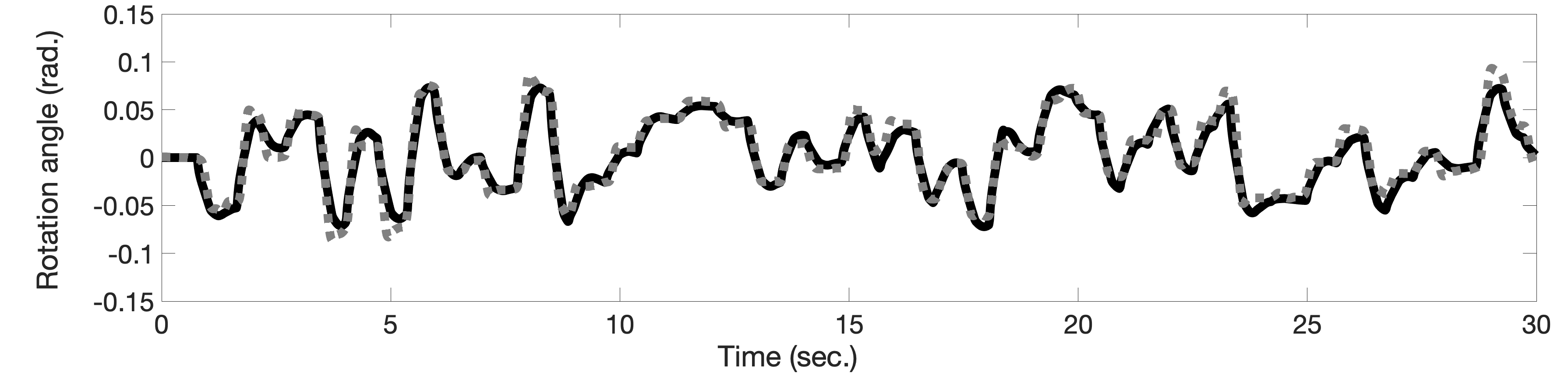}}
\qquad
\subfloat[Subject 8]{\label{fig:08}\includegraphics[width=0.95\textwidth]{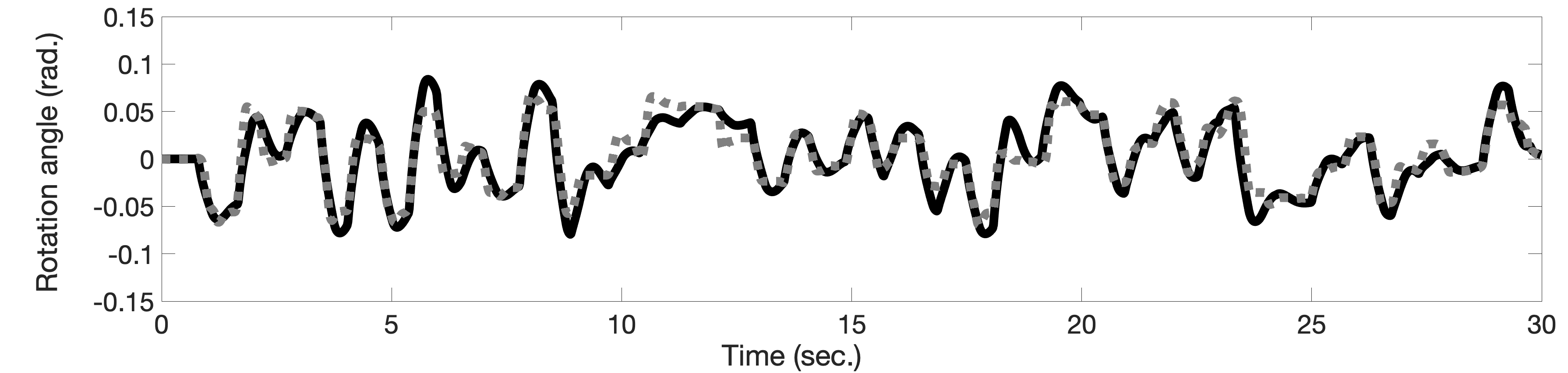}}
\qquad
\subfloat[Subject 9]{\label{fig:09}\includegraphics[width=0.95\textwidth]{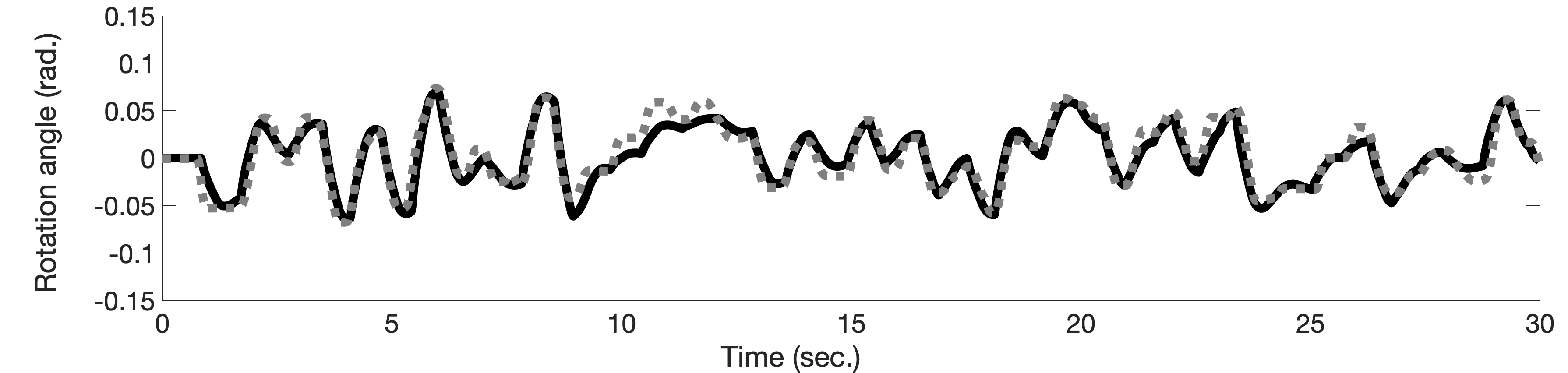}}
\qquad
\subfloat[Subject 10]{\label{fig:10}\includegraphics[width=0.95\textwidth]{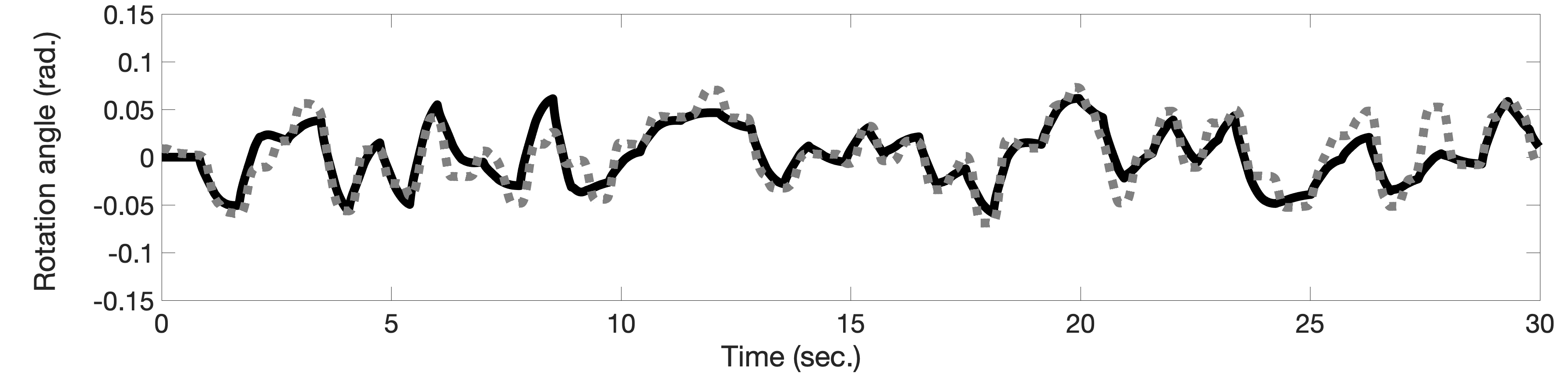}}
\caption{The curve fitting with 10 experimental cases (continued). Solid lines represent estimated responses from the fitted models and dotted lines represent measured responses.}
\label{fig:cont}
\end{figure*}




\begin{table}[t]
\centering
\begin{tabular}{c c }
\hline
\hline
\textbf{Subject}& \textbf{Improvement ($\%$)}\\ \hline
1 & 69.25 \\
2 & 65.79 \\ 
3 & 74.83 \\ 
4 & 89.36 \\ 
5 & 87.44 \\ 
6 & 65.02 \\ 
7 & 53.62 \\
8 & 77.70 \\ 
9 & 63.44 \\ 
10& 67.62 \\ \hline
Avg. & 71.41 \\ \hline \hline
\end{tabular}
\caption{Improvement on the variance of estimated parameters in an experiment study. The given values are $(1-\frac{{\sigma}^{2,\text{Lasso}}}{{\sigma}^{2,\text{All}}}) \times 100$. ${\sigma}^{2,\text{Lasso}}$ is the variance of estimated parameters obtained from our method, and ${\sigma}^{2,\text{All}}$ is the variance of estimated parameters without regularization.}
\label{sim_2}
\end{table}
\subsection{The Experiment}\label{data}
As shown in Fig.~\ref{fig2}, each  subject  wears  a helmet attached with two string potentiometers measuring the axial rotation of the head. Subjects rotated  their heads about the vertical axis to track the command signals on the display. The command signal $r(t)$ is a pseudorandom sequence of steps with random step durations and amplitudes. The angle of the signal is bounded between  $\pm4^{\circ}$. The output signal $y(t)$ is the head rotation angle. Each subject performed three 30-second trials, and the sampling rate was 60 Hz~\cite{ramadan2018selecting}.
\subsection{Simulation study}
In this section, we analyze the bias and variance of estimated parameters from a simulation study with the known true parameters for comparison. First, we generate the simulated data $(\bm x, \bm y)$ where $\bm x \in \mathbb{R}^{1800\times 1}$ and $\bm y \in \mathbb{R}^{1800\times 1}$, which are the input and  observation vectors, respectively.
Additionally, we obtain 20 sets of nominal parameter values from preliminary estimation performed 20 times over all three trials for each subject. Finally, we evaluate our method by setting the mean of 20 sets of nominal parameter values as a set of typical values.
In Fig.~\ref{bv}, as compared to the nonlinear least squares estimator without ${L}_{1}$-regularization, except for $\tau_{c}$, the biases of the other parameters were decreased by $28.0\%$ on average. In addition, the variances of all estimated parameters were decreased by $96.1\%$ on average.
\sloppy
\subsection{In vivo experimental study} \label{ex}
In this section, the variances of estimated parameters from our method are compared with the result of the standard simplex-based optimization in \cite{ramadan2018selecting}. All parameters are pushed further toward the mean of nominal parameter values obtained from preliminary estimation as the regularization hyperparameter increases. The regularization hyperparameter increases until 5 sensitive parameters are selected. After selecting a sensitive parameter subset for each subject, we select the most frequent subset among all subjects for the fair comparison with \cite{ramadan2018selecting}. The Lasso in \cite{ramadan2018selecting} selected 5 parameters of $ [{K}_{vcr} \ {K}_{ccr} \  \tau_{1A} \  \tau_{c} \ \tau_{CNS2} ]$ as the most frequent subset of sensitive parameters, and the subset of $ [ {K}_{ccr} \  \tau \ \tau_{1A} \  \tau_{c} \ \tau_{CNS2} ]$ was selected by our method. As a result, 4 out of 5 sensitive parameters $ [ {K}_{ccr} \  \tau_{1A} \  \tau_{c} \ \tau_{CNS2}]$ are selected by both methods. 

Second, we evaluate our method based on the goodness of fit measured by variance accounted for (VAF). All values are given as mean $\pm$ standard deviation across subjects. The goodness of fit of our method ($\text{VAF} = 82.5\% \pm 8.3\%$) over 10 subjects is almost equal to that of the Lasso ($\text{VAF} = 83.3\% \pm 7.3\%$ ) in \cite{ramadan2018selecting}. Without ${L}_{1}$-regularization, $\text{VAF} = 84.9\% \pm 0.1\%$ over all subjects. In Fig.~\ref{fig:cont}, for all subjects, the estimated responses are almost same as the measured responses, and the estimated responses are smoother than the measured responses.

Third, as shown in Table~\ref{sim_2}, the variance of estimated parameters is reduced by 71.4$\%$ on average across parameters and subjects as compared to those of estimated parameters without regularization.

Finally, we compute the average computation time using the ``timeit'' function from   $MATLAB$ (The MathWorks Inc., Natick, MA, U.S.A.). The average computation time of our method for a subject is 54 times faster than that of the Lasso with the standard simplex-based optimization in \cite{ramadan2018selecting}. In particular, the average computation time of our method across subjects is 5.6 seconds per trial, and that of the Lasso in \cite{ramadan2018selecting} is 302.0 seconds per trial.

\section{Discussion}
\sloppy
We provided consistency and oracle properties (i.e., convergence to the correct sparsity and asymptotic normality) for a nonlinear regression approach with a generalized penalty function. As a result, we proved the existence of a local minimizer and the convergence to the sparse unknown parameters for the penalized nonlinear least squares estimator. It is important to note that for the first time, we have proved convergence properties of the penalized nonlinear least squares estimator, as compared to previous studies \cite{johnson2008penalized,wu1981asymptotic,fan2001scad}.
 
\sloppy
In the simulation study, we showed that the bias and variance of estimated parameters of our method were decreased as compared to those of estimated parameters without ${L}_{1}$-regularization. When we set the mean of nominal parameter values as the typical values for non-selected estimates, the variance significantly was decreased. In addition, although the ${L}_{1}$-regularization is known to induce biased estimates, the bias with our method also slightly was decreased except for one parameter with the increased bias. The reason might be that our method pushes all parameters toward the mean of nominal parameter values obtained from 20 preliminary estimation. Therefore, if we set the appropriate  values as the typical values, we could achieve lower values of bias and  variance errors.

In the experimental study, our proposed method was compared with the Lasso in \cite{ramadan2018selecting} using three performance criteria.
First, we confirmed that the proposed method simultaneously selected and estimated sensitive parameters in the nonlinear model.
As a result, five parameters were selected and estimated in our method, and the remaining parameters were fixed to the mean of nominal parameters obtained from preliminary estimation. With our method, 4 out of 5 sensitive parameters were the same as those selected in \cite{ramadan2018selecting}. This result showed that our method behaved similarly in sensitive parameter selection by \cite{ramadan2018selecting} using the Fisher information matrix. Selected sensitive parameters may vary slightly depending on the performance of the optimizer and the condition of initial points. However, they have not changed much over repeated randomized realizations. In addition, we presented VAF to quantitatively evaluate the goodness of fit of the estimated model. From the standard nonlinear least squares problem without ${L}_{1}$-regularization, VAF was about 84.9$\%$, and our method achieved  about 82.5$\%$. Hence, the goodness of fit of the model estimated from our method was similar to that of model estimated from the standard nonlinear  least squares problem without ${L}_{1}$-regularization although only 5 selected parameters were used for estimation in our method. Moreover, as shown in Fig.~\ref{fig:cont}, most curve fitting errors occur around the peak points, which shows the limitation of the presented dynamics models that do not perfectly reflect the real physiological head-neck control processes. This could be due to the switching of human control strategies with sudden changes in head-neck orientations~\cite{chen2002modeling}.

Next, the model identifiability was improved by sensitive parameter selection from our method. The variance of estimated 12 parameters is reduced by $71.1\%$ on average. In general, when the number of parameters to be estimated is large with limited  data, the model has poor (or lack of) identifiability. Therefore,  through our method with key parameter selection, i.e., the nonlinear least squares problem with $L_1$-regularization on the deviation of parameters from the mean of the nominal parameter values, the uniqueness of the estimated solution can be ensured even for an original problem with a lack of identifiability due to limited data. 

Finally, the average computation time of our method was 54 times faster than that of the Lasso with the brute force optimization algorithm in \cite{ramadan2018selecting}. Our method reduced the computation time by eliminating large inverse matrix computation by modifying a Jacobian formulation as a minimization formulation.
\sloppy
\section{Conclusions}
In this paper, we tackled a parameter estimation problem with limited data by formulating it as a nonlinear least squares estimator with $L_1$-regularization. 

We effectively  improved the model identifiability by applying the Lasso to the nonlinear least squares problem. As asymptotic results, we provided consistency and oracle properties for a nonlinear regression approach with a generalized penalty function. Based on these results, we proposed a novel solution to our problem by solving   the nonlinear least squares problem with ${L}_{1}$-regularization on the deviation of parameters from the  nominal  values in order to simultaneously select and estimate model parameters. 

From simulation and experimental studies, we successfully  demonstrated   that the proposed method simultaneously selected and estimated sensitive parameters, improved the model identifiability by reducing the variance of estimated parameters and took a much shorter computation time than that of the Lasso in \cite{ramadan2018selecting}. 

Future work would be to apply our method to other clinical patient-specific calibration of the disease models \cite{do2018prediction, zhang2019patient}.

\begin{acknowledgement}
This work was supported by the Mid-career Research Programs through the National Research Foundation of Korea (NRF) funded by the Ministry of Science and ICT (NRF-2018R1A2B6008063, 2019R1A2C1002213). This publication was made possible by grant number U19AT006057 from the National Center for Complementary and Integrative Health (NCCIH) at the National Institutes of Health. Its contents are solely the responsibility of the authors and do not necessarily represent the official views of NCCIH.
The research of Wei-Ying Wu was supported by Ministry of Science and Technology of Taiwan under grants (MOST 107-2118-M-259-001-). 
\end{acknowledgement}

\section*{Conflict of interest}
The authors declare that they have no conflict of interest.

\begin{appendices}
\section{Proof of Lemma~\ref{lem:Consistency}}\label{A_1}
	We first find the lower bound of  $\bm Q_n(\bm\theta_0+n^{-1/2} \bm v)-\bm Q_n(\bm\theta_0)$. 
	\begin{align*}
	& \bm Q_n(\bm \theta_0+n^{-1/2} \bm v)-\bm Q_n(\bm\theta_0)\\
	&=\bm S_n(\bm \theta_0+n^{-1/2} \bm v)-\bm S_n(\bm\theta_0)\\
	&~~~+n(\sum^{p}_{k=1}p_{\lambda_n}(| \theta_{0k}+n^{-1/2} v_k|)-\sum^{p}_{k=1}p_{\lambda_n}(|\theta_{0k}|))\\
	&= n^{-1/2}\left(\nabla \bm S_n(\bm\theta_0)\right)^{T}\bm{v}
	+\frac{1}{2}n^{-1}\bm v^{T}\nabla^2 \bm S_n(\bm\theta^{*}) \bm v\\
					    \end{align*}
  	\begin{align*}
	& ~~~+n\left(\sum^{s}_{k=1}p_{\lambda_n}(|\theta_{0k}+n^{-1/2} v_k|)-\sum^{s}_{k=1}p_{\lambda_n}(|\theta_{0k}|)\right) \\
	&~~~+n\sum^{p}_{k=s+1}p_{\lambda_n}(|\theta_{0k}+n^{-1/2} v_k|), \text{where}\\
	& ~~~ \hbox{$\bm\theta^{*}=(\theta^{*}_1, \theta^{*}_2, \ldots, \theta^{*}_p)$ lies between $\bm\theta_0+n^{-1/2}\bm v$ and $\bm\theta_0$},\\
	&= n^{-1/2}\left(\nabla \bm S_n(\bm\theta_0)\right)^{T} \bm{v}+\frac{1}{2}n^{-1} v^{T}\nabla^2 \bm S_n(\bm\theta^{*}) v\\ &~~+n\left(\sum^{s}_{k=1}q_{\lambda_n}(|\theta^{*}_{0k}|)\text{sgn}(\theta^{*}_{0k})n^{-1/2} v_k\right)\\
	&~~+n\sum^{p}_{k=s+1}p_{\lambda_n}(|\theta_k+n^{-1/2} v|)\\
	&\geq n^{-1/2} (\nabla \bm S_n(\bm\theta_0))^{T} \bm{v}+\frac{1}{2}n^{-1}v^{T}\nabla^2 \bm S_n(\bm\theta^{*}) (v)\\ &~~+n\left(\sum^{s}_{k=1}q_{\lambda_n}(|\theta^{*}_{0k}|)\text{sgn}(\theta^{*}_{0k})n^{-1/2}v_k\right)\\
	&:=\mathbb{A}+ \mathbb{B}+ \mathbb{C}.
	\end{align*}
	For the sake of simplicity, \\
	we define ${\bm{d}}=(d_1(\bm\theta, \bm\theta^{\prime}),...,d_n(\bm\theta, \bm\theta^{\prime}))$, where $d_i(\bm\theta, \bm\theta^{\prime})=f({\bm{x}}_i;
	\bm\theta)-f({\bm{x}}_i;\bm\theta^{\prime})$. 	
	For the term $\mathbb{A}$,
	\begin{eqnarray*}
		&&n^{-1/2}(\nabla \bm S_n(\bm\theta_0))^{T}\bm{v}\\
		&&~~~=-2n^{-1/2}\bm{v}^{T}\bm{\dot{F}}(\bm\theta_0)\bm{\epsilon},\\
		&&~~~=-2\bm{v}^{T}\left[\bm{F}_n^{1/2}\left(\bm{\dot{F}(\theta_{0})}^{T}\bm{\dot{F}(\theta_{0})}\right)^{-1/2}\bm{\bm{\dot{F}}}(\bm\theta_0)^{T}\bm{\epsilon}\right], \\
		&& ~~~\hbox{where } \bm{F}_0 =\frac{1}{n}\bm{\dot{F}(\theta_{0})}^{T}\bm{\dot{F}(\theta_{0})}.
	\end{eqnarray*} 
	By Assumption~\ref{eq:nonlinear:assumption}, $\bm{F}_0^{1/2}\ \rightarrow \ \Gamma^{1/2}$ and we claim that,
	$$\left(\bm{\dot{F}}(\bm\theta_0)^{T} {\bm{\dot{F}}}(\bm\theta_0) \right)^{-1/2}{\bm{\dot{F}}}(\bm\theta_0)^{T}{\bm \epsilon}~\stackrel{d}{\longrightarrow} N(0, \sigma^{2}I). $$
	The claim follows from the lemma 2.1 in \cite{huber1973robust}. The condition for the lemma in our setting is
	$$\left\| \bm{\dot{F}}(\bm\theta_0) \left(\bm{\dot{F}}(\bm\theta_0)^{T} {\bm{\dot{F}}}(\bm\theta_0)\right) ^{-1/2}\right\|_{\infty}\rightarrow 0 $$
	where $\|A\|_{\infty}$ denote the maximum absolute value of all elements of matrix $A$. Since $\bm{\dot{F}}(\bm\theta_0)$ is an $n\times p$ matrix and $\left(\bm{\dot{F}}(\bm\theta_0)^{T}{\bm{\dot{F}}}(\bm\theta_0)\right)^{-1/2}$ is a $p\times p$ matrix,
	
	\begin{align*}
	& \left\| \bm{\dot{F}}(\bm\theta_0) \left(\bm{\dot{F}}(\bm\theta_0)^{T} {\bm{\dot{F}}}(\bm\theta_0)\right) ^{-1/2}\right\|_{\infty}  \\ & \leq  p\left\|\left(\bm{\dot{F}}(\bm\theta_0)^{T} {\bm{\dot{F}}}(\bm\theta_0)\right) ^{-1/2}  \right\|_{\infty}\left\|\bm{\dot{F}}(\bm\theta_0) \right\|_{\infty}\\
	 & = pn^{-1/2}\left\|\bm{F}_0^{-1/2}  \right\|_{\infty}\left\|\bm{\dot{F}}(\bm\theta_0) \right\|_{\infty}\\
	 & = O\left(\frac{p}{\sqrt{n}}\right).
	\end{align*}
	The first and second conditions from Assumption~\ref{eq:nonlinear:assumption} imply the last equality.
	Therefore, we have
	\begin{equation}
	\begin{aligned}
		-2\bm{v}^{T}\left[\bm{F}_0^{1/2}\left(\bm{\dot{F}(\theta_{0})}^{T}\bm{\dot{F}(\theta_{0})}\right)^{-1/2}\bm{\bm{\dot{F}}}(\bm\theta_0)^{T}\bm{\epsilon}\right]\\
		 ~\stackrel{d}{\longrightarrow}~N(0,4\sigma^2 \bm v^{T}\Gamma \bm v).
	\end{aligned}
	\end{equation}
	For the term $\mathbb{B}$,
	\begin{align*}
	&\frac{1}{2}n^{-1} \bm v^{T}\nabla^2 \bm S_n(\bm\theta^{*})\bm v\\
	&=n^{-1}\bm v^{T}\left\{\left({\bm{\dot{F}}}(\bm\theta^{*})^{T} {\bm{\dot{F}}}(\bm\theta^{*}) \right)+{\bm{{\ddot{F}}}}(\bm\theta^{*})^{T}(I\otimes(\bm d-\bm \epsilon))\right\}\bm v\\
	&=\bm v^{T}\left[\frac{\left({\bm{\dot{F}}}(\bm{\theta_0})^{T} {\bm{\dot{F}}}(\bm{\theta_0}) \right)}{n} \bm Z^{-1}_n \bm v\right], \hbox{where}\\
	&~~~\bm Z_n=\left\{\left({\bm{\dot{F}}}(\bm\theta^{*})^{T} {\bm{\dot{F}}}(\bm\theta^{*}) \right)+{\bm{{\ddot{F}}}}(\bm\theta^{*})^{T}(I\otimes(\bm d-\bm\epsilon)\right\}^{-1}n\bm{F}_0.
	\end{align*}
	If we show $\bm{Z_n}^{-1}\stackrel{p}{\rightarrow} I$, with Assumption~\ref{eq:nonlinear:assumption}, we obtain 
	\begin{equation}
	\label{eq:AB}
	\begin{aligned}
	\mathbb{A}+\mathbb{B}~ \stackrel{d}{\longrightarrow}~ N( \bm v^{T}\Gamma \bm v,4\sigma^2 \bm v^{T}\Gamma \bm v).
	\end{aligned}
	\end{equation}
 	$\bm{Z_n}^{-1}$ can be rewritten as,
	\begin{align*}
		\bm{Z_n^{-1}}&=\left(n\bm{F}_0\right)^{-1}\left\{\left({\bm{\dot{F}}}(\bm\theta^{*})^{T} {\bm{\dot{F}}}(\bm\theta^{*}) \right)+\bm{{\ddot{F}}}(\bm\theta^{*})^{T}(I\otimes(\bm d-\bm\epsilon)\right\}\\
		&=\left(n\bm{F}_0\right)^{-1}\left({\bm{\dot{F}}}(\bm\theta^{*})^{T} {\bm{\dot{F}}}(\bm\theta^{*}) \right)+\left(n\bm{F}_0\right)^{-1}\bm{{\ddot{F}}}(\bm\theta^{*})^{T}(I\otimes \bm{d}) \\
		& - \left(n\bm{F}_0\right)^{-1}\bm{{\ddot{F}}}(\bm\theta^{*})^{T}(I\otimes \bm{\epsilon}).
	\end{align*}
	By Assumption~\ref{eq:nonlinear:assumption}, the first term converges to $I$ almost surely. By the conditions 1, 2, and 4 of Assumption~\ref{eq:nonlinear:assumption} and Cauchy-Schwarz inequality, the second term converges to zero almost surely. For the last term, it is enough to show that
	\begin{equation}
	\begin{aligned}
	\frac{1}{n}\bm{f_{ks}}(\bm{\theta})^{T}\bm{\epsilon}\ \stackrel{p}{\longrightarrow} 0
	\end{aligned}
	\end{equation}
	uniformly on $S-\{\theta \in \Theta : \|\theta-\theta_{0}\| \leq \delta \}$ with probability 1, because of the conditions 1 and 2 of Assumption~\ref{eq:nonlinear:assumption}. Then, by the condition 4 of Assumption~\ref{eq:nonlinear:assumption}, (3.17) can be shown in a manner similar to that of \cite{wu1981asymptotic} 
	For the term $\mathbb{C}$, by Assumption~\ref{eq:penalty:assumption} for a fixed $s<\infty$, 
	\begin{equation}
	\label{eq:C}
	\begin{aligned}
	\left(n^{1/2}\sum^{s}_{k=1}q_{\lambda_n}(|\theta^{*}_{0k}|)\text{sgn}(\theta^{*}_{0k})v_k\right)~\longrightarrow~ 0.
	\end{aligned}
	\end{equation}
	Thus, combined~(\ref{eq:AB}) with~(\ref{eq:C}), we have 
	$$\mathbb{A}+\mathbb{B}+\mathbb{C}~\stackrel{d}{\longrightarrow}~ N( \bm v^{T}\Gamma \bm v,4\sigma^2 \bm v^{T}\Gamma \bm v),$$
which leads to the desired result for large enough $C$.

\section{Proof of Theorem~\ref{thm:Consistency}}\label{A_2}

    By Lemma~\ref{lem:Consistency} and the continuity of $\bm Q_n(\cdot)$, we obtain Theorem~\ref{thm:Consistency}.

\section{Proof of Theorem~\ref{thm:sparsity}}\label{A_3}

    \noindent Proof of (i)\\
	First, we break $M_k$ into two sets: 	
	\begin{align*}
	M_k&=\left\{\omega:\hat{\theta}_k\neq 0, |\hat{\theta}_k|\geq C n^{-1/2}\right\} \\
	&~~~~+\left\{\omega:\hat{\theta}_k\neq 0, |\hat{\theta}_k|< C n^{-1/2}\right\}\\
	&=: E_n+F_n.
	\end{align*}
	Then, it is enough to show for any $\epsilon > 0$, $P(E_n) <\epsilon/2 $ and $P(F_n) < \epsilon/2$. For any $\epsilon >0$, we can show $P(E_n)<\epsilon/2$ for large enough $n$ because of the consistency.
	
    To verify $P(F_n)<\epsilon/2$ for large enough $n$, we first show $n^{1/2}q_{\lambda_n}(|\hat\theta_k|)=O_p(1)$ on the set
	$F_n$. Note that 	
	\begin{eqnarray*}	
		&&n^{-1/2}\nabla \bm S_n(\bm\theta)-n^{-1/2}\nabla \bm S_n(\bm\theta_0)\\
		&&= n^{-1/2}\nabla^2 \bm S_n(\bm\theta^{**})(\bm\theta-\bm\theta_0)\\
		&&=n^{1/2}\frac{\left({\bm{\dot{F}}}(\bm\theta^{**})^{T} {\bm{\bm\dot{F}}}(\bm\theta^{**}) \right)}{n}\bm Z^{-1}_n(\bm\theta-\bm\theta_0)=O_p(1),
	\end{eqnarray*}
	where $\bm\theta^{**}$ lies between $\bm\theta$ and $\bm\theta_0$. Since $\frac{1}{n}{\bm{\dot{F}}}(\bm\theta^{**})^{T} {\bm{\dot{F}}}(\bm\theta^{**})  ~\stackrel{p}{\longrightarrow}~ \Gamma$, $\bm Z^{-1}_n ~\stackrel{p}{\longrightarrow}~ I$ and $\|\bm\theta-\bm\theta_0\|=O_p(n^{-1/2}).$ Thus, we have
	\begin{equation}\label{eq:firstderivative}
	\sup_{\|\theta-\theta_0\|\leq Cn^{-1/2}}\|n^{-1/2}\nabla \bm S_n(\bm\theta)-n^{-1/2}\nabla \bm S_n(\bm\theta_0)\|=O_p(1).
	\end{equation}
	
	Combining~(\ref{eq:firstderivative}) with $\|n^{-1/2}\nabla S_n(\theta_0)\|=O_p(1)$, we have
	\begin{equation}\label{eq:Sn}
	\|n^{-1/2}\nabla \bm S_n(\bm \theta)\|=O_p(1)
	\end{equation}
	for $\bm\theta$ which satisfies $\|\bm\theta-\bm\theta_0\|\leq Cn^{-1/2}.$
	Since $\hat{\bm\theta}$ is the local minimizer of $\bm Q_n(\bm\theta)$ with the root-n-consistent, we attain
	\begin{equation}{\label{eq:penalty}}
	n^{1/2}q_{\lambda_n}(|\hat{\theta}_k|)=O_p(1)
	\end{equation}
	from 
	\begin{align*}
	&n^{-1/2}\frac{\partial \bm Q_n(\bm \theta)}{\partial \theta_k}\Big|_{\bm\theta=\hat{\bm\theta}}\\ 
	&~~~~~~~~~~~~~~~=n^{-1/2}\frac{\partial \bm S_n(\bm \theta)}{\partial\theta_k}\Big|_{\bm\theta=\hat{\bm\theta}}+n^{1/2}q_{\lambda_n}(|\hat{\theta_k}|)\text{sgn}(\hat\theta_k).
	\end{align*}
	Therefore, there exists $M^{\prime}$ such that,  for large enough $n$,
	$$P\{\omega:\hat{\theta}_k\neq 0, |\hat{\theta}_k|<C n^{-1/2},  n^{1/2}q_{\lambda_n}(|\hat{\theta}_k|)>M^{\prime}\}<\epsilon/2.$$ In addition, by the second assumption of Assumption~\ref{eq:penalty:assumption}, 
	\begin{align*}
		&\{\omega:\hat{\theta}_k\neq 0, |\hat{\theta}_k|<C n^{-1/2},  n^{1/2}q_{\lambda_n}(|\hat{\theta}_k|)>M^{\prime}\}\\
		&~~~~~~~~~~~~~~~~~~~~~~~~=\{\omega:\hat{\theta}_k\neq 0, |\hat{\theta}_k|<C n^{-1/2}\}
	\end{align*} for the large enough $n$, which leads to $P(F_n)<\epsilon/2$ for large enough $n$. \\
	
	\noindent Proof of (ii)\\
	By the Taylor expansion,
	\begin{eqnarray*}	
		n^{-1/2}\nabla \bm Q_n(\hat{\bm\theta})&=&n^{-1/2}\nabla \bm S_n(\hat{\bm\theta})+n^{-1/2}{\bm{q}}_{\lambda_n}(\hat{\bm\theta})\cdot \text{sgn}(\hat{\bm\theta})\\
		&= &n^{-1/2}\left(\nabla \bm S_n(\bm\theta_0)+\nabla^2 \bm S_n(\bm\theta^{**})(\hat{\bm\theta}-\bm\theta_0)\right)\\
		&&~+n^{1/2}{\bm{q}}_{\lambda_n}(\hat{\bm\theta})\cdot \text{sgn}(\hat{\bm\theta})
	\end{eqnarray*} where ${\bm{q}}_{\lambda_n}(\hat{\bm\theta}) \cdot \text{sgn}(\hat{\bm\theta})=\left(q_{\lambda_n}(|\hat{\theta}_1|)\text{sgn}(\hat\theta_1),..,q_{\lambda_n}(|\hat{\theta}_p|)\times\right.$ $\left. \text{sgn}(\hat\theta_p)\right)^{T}.$
	Since $\hat{\bm\theta}$ is the local minimizer of $\bm Q_n({\bm\theta})$,
	$\nabla \bm Q_n(\hat{\bm\theta})=0$ so that 
	\begin{align*}
	n^{-1/2}(-\nabla \bm S_n({\bm\theta_0}))= & n^{-1}\nabla^2 S_n(\bm\theta^{**})\left(n^{1/2}(\hat{\bm\theta}-\bm\theta_0)\right)\\ & +n^{1/2}{\bm{q}}_{\lambda_n}(\hat{\bm\theta})\cdot \text{sgn}(\hat{\bm\theta}).
	\end{align*}
	Finally,$$n^{1/2}2\Gamma_{11}({\hat{\bf{\bm\theta}}}_{11}-\bm\theta_{01}+(2\Gamma_{11})^{-1}b_n)~ \stackrel{d}{\longrightarrow}~ N(0,4\Gamma_{11}\sigma^2),$$ because $n^{-1}\nabla^2 \bm S_n(\bm\theta^{**}) \stackrel{p}{\longrightarrow} 2\Gamma$, $n^{-1/2}(-\nabla \bm S_n({\bm\theta_0}))  \stackrel{d}{\longrightarrow}$ $N({\bm{0}},4\Gamma\sigma^2)$ and  the consistency of $\hat{\bm\theta}$.

\section{Nonlinear least squares estimator with ${L}_{1}$-regularization.}\label{ours}
In order to apply the Lasso to the nonlinear least squares problem, we reformulate the Levenberg-Marquardt (LM) optimization algorithm as a linear least squares problem as follow.
\begin{equation}\label{ourmethod}
\begin{aligned}
    {\hat{\tilde{\bm\theta}}}^{j+1} &= \min\limits_{{\tilde{\bm\theta}}} \| \bm\Lambda {\tilde{\bm\theta}}^{j+1}-\bm\Lambda{\tilde{\bm\theta}}^{j} + \boldsymbol{J}^{T}{\tilde{\bm r}}({\tilde{\bm\theta}}^{j}) \|^{2}_{2}\\
    &\quad \ s.t.\  \sum^{p}_{k=1} |\tilde{\theta}^{j+1}_{k}| \leq {T} _{\theta},
\end{aligned}
\end{equation}
where ${\tilde{\bm\theta}}^{j+1} = {\tilde{\bm\theta}}^{j} + \Delta\bm\theta , \Delta\bm\theta = -\bm\Lambda^{-1}\bm{J}^{T}{\tilde{\bm r}}({\tilde{\bm\theta}}^{j}),\bm\Lambda=(\bm{J}^{T}\bm{J} + \mu \text{diag}  (\bm{J}^{T}\bm{J})) $.
The residual vector ${\tilde{\bm r}}({\tilde{\bm\theta}}^{j})=\bm{r}({\tilde{\bm\theta}}^{j}+{\bar{\bm\theta}})=\bm{r}(\bm{\theta}^j)=\bm y-{f}(\bm{X};\bm\theta^{j})$. ${T}_{\theta}$ is the regularization hyperparameter. The sum of the deviation of parameters from the mean of the nominal parameter values is less than or equal to the regularization hyperparameter ${T}_{\theta}$.
 The damping factor $\mu$ affects the efficiency and the convergence stability~\cite{cui2017new}. $\bm J$ is the Jacobian matrix which consists of all first-order partial derivatives of the residual vector with respect to the parameters, evaluated for ${\tilde{\bm\theta}} = {\tilde{\bm\theta}}^{j}$ as follows.
\begin{equation*}
\bm J := \frac{\partial {\tilde{\bm r}}({\tilde{\bm\theta}}^{j})}{\partial {\tilde{\bm\theta}}^{j}}
\end{equation*}
In addition, as compared with the conventional Lasso fixing insensitive parameters to 0, our method simultaneously selects and estimates only sensitive parameters while fixing insensitive parameters onto the mean of the nominal parameter values ${\bar{\bm\theta}}$.
\end{appendices}

\begin{table}[!t]\normalsize
  \centering
  \caption{The algorithm of sensitive parameter selection using our method}
  \label{alg:lasso}
  \begin{tabular}{| l l |}
    \hline
    \textbf{Input:}  &
    \begin{minipage}{0.37\textwidth}
      \vspace{4pt} (1) Experimental Data and Dynamical Model\\
      (2) Vector of Normalized Values (${\bar {\bm\Phi}}$), where the Mean of Nominal Parameter Values is ${\bar{ \bm\theta}}$ \\
      (3) Desired Optimal Number of Sensitive Parameters (${{n}^{*}}$)\vspace{2pt}\\
      (4) The Regularization Hyperparameter ${T}_{\theta}$
    \end{minipage} \\
    \textbf{Output:} &
    \begin{minipage}{0.37\textwidth}
      \vspace{2pt} (1) A Subset of the Sensitive Parameters ($\hat { \tilde  { \bm\theta}}$)  \vspace{2pt}
    \end{minipage} \\
    \hline \multicolumn{2}{| l |}{
      \begin{minipage}{0.37\textwidth}
        \vspace{4pt}
        \begin{algorithmic}[1]
          \STATE $\text{NumParams} = 0$
          \STATE ${T}_{\theta}= 1.0$ 
          \WHILE {$\text{NumParams} \hspace{5pt}!= {{n}^{*}}$} 
          \REPEAT
          \STATE Solve (\ref{ourmethod}) in Appendix D
          \UNTIL {$ \tilde  { \bm\Phi} $ convergences}
          \FOR {$i = 1:n$} 
          \IF {$\hat { \tilde  { \bm\Phi}}(i) > 0.001$}
          \STATE NumParams = NumParams + 1
          \ENDIF
          \ENDFOR
          \IF{$\text{NumParams} \hspace{5pt}!= {{n}^{*}}$}
          \STATE ${T}_{\theta} = {T}_{\theta} - \frac{\text{NumParams}-{{n}^{*}}}{i+{n}^{*}}$
          \ENDIF
          \ENDWHILE \\
          ${\hat{\tilde{\bm\theta}}}$, where a subset of the normalized sensitive parameters is $ \hat { \tilde  { \bm\Phi}}$
          \vspace{4pt}
        \end{algorithmic}
      \end{minipage}
    } \\
    \hline
  \end{tabular}
\end{table}


%
%

\bibliographystyle{spmpsci}      
\bibliography{reference}   


\end{document}